\documentstyle[12pt]{article}
\normalsize
\def\simless{\mathbin{\lower 3pt\hbox{$\rlap{\raise 5pt\hbox{$\char'074$}}
\mathchar"7218$}}}
\def\simgreat{\mathbin{\lower 3pt\hbox{$\rlap{\raise 5pt \hbox{$\char'076$}}
\mathchar"7218$}}}

\def\ch{\@startsection{section}{1}{\z@}{-3ex plus-1ex minus-.2ex}%
        {2ex plus.2ex}{\large\sc}}

\def\; \lapp \;{\raisebox{-.4ex}{\rlap{$\sim$}} \raisebox{.4ex}{$<$}}

\def\con{\ifmmode \hbox{\bf*} \else{\bf*}\fi}   
\def\scon{\ifmmode \hbox{\footnotesize\rm\bf*} \else{\footnotesize\rm\bf*}\fi}

\def\0#1{\relax\ifmmode\mathaccent"7017{#1}
        \else\accent23#1\relax\fi}              

\def\eslash{\not{\hbox{\kern-2pt $E$}}}

\catcode`@=12

\begin{document}
\hoffset=0.4cm
\voffset=-1truecm
\normalsize


\begin{titlepage}
\begin{flushright}
DFPD 93/TH/26\\
UTS-DFT-93-9\\
SISSA 93/50-A\\
hep-ph 9304267\\
\end{flushright}
\vspace{24pt}
\centerline{\Large {\bf Spontaneous CP Violation and Baryogenesis}}
\vskip 0.1 cm
\centerline{\Large {\bf in the Minimal Supersymmetric Standard Model}}
\vspace{24pt}
\begin{center}
{\large\bf D. Comelli$^{a,b,}$\footnote{Email: comelli@mvxtst.ts.infn.it}
, M. Pietroni$^{c,d,}$\footnote{Email: pietroni@mvxpd5.pd.infn.it}
 and A. Riotto$^{d,e,}$\footnote{Email:riotto@tsmi19.sissa.it}}
\end{center}
\vskip 0.2 cm
{\footnotesize
\centerline{\it $^{(a)}$Dipartimento di Fisica Teorica Universit\`a di
Trieste,}
\centerline{\it Strada Costiera 11, 34014 Miramare, Trieste, Italy}
\vskip 0.2 cm
\centerline{\it $^{(b)}$ Istituto Nazionale di Fisica Nucleare,}
\centerline{\it Sezione di Trieste, 34014 Trieste, Italy}
\vskip 0.2 cm
\centerline{\it $^{(c)}$Dipartimento di Fisica Universit\`a di Padova,}
\centerline{\it Via Marzolo 8, 35100 Padua, Italy.}
\vskip 0.2 cm
\centerline{\it $^{(d)}$Istituto Nazionale di Fisica Nucleare,}
\centerline{\it Sezione di Padova, 35100 Padua, Italy.}
\vskip 0.2 cm
\centerline{\it $^{(e)}$International School for Advanced Studies,
SISSA-ISAS}
\centerline{\it Strada Costiera 11, 34014 Miramare, Trieste, Italy}}
\vskip 0.5 cm
\centerline{\large\bf Abstract}
\vskip 0.2 cm
\baselineskip=15pt
We investigate the effects of the spontaneous CP violation at finite
temperature in the Minimal Supersymmetric Standard Model on the
baryogenesis at the weak scale. After a brief discussion of the case in
which the electroweak phase transition is of the second order, we study
in details the baryogenesis scenario when the transition proceeds via
bubble nucleation. We show that the space-time
dependent phase for the Higgs vacuum expectation values coming from
the spontaneous CP violation can give rise to a efficient generation
of baryon number inside the bubble walls if the superpotential parameters and
the soft supersymmetry breaking ones are complex. However we
find that in order to get the observed value for the baryon asymmetry of
the universe the phases of such parameters can be as small as $10^{-5}$,
giving rise to an electron dipole moment of
the neutron well below the current experimental limit. Moreover a light
Higgs pseudoscalar is needed, and an upper bound on its mass is
obtained.
\end{titlepage}
\baselineskip=14pt

{\large {\bf 1. Introduction}}

\vskip 0.8 cm

The intriguing suggestion \cite{dim,kuz} that the three Sakharov's conditions
 \cite{sak} for the generation of the Baryon Asymmetry in the Universe (BAU)
might be fulfilled at the Electro Weak Phase Transition (EWPT) has stimulated
a lot of theoretical effort in the last few years \cite{dolgov}.
Nevertheless, there
are still uncertainties concerning each of the three
requirements, namely the departure from the thermodynamic equilibrium,
the presence of baryon number (B) violating processes, and CP violation.

The out of equilibrium condition can be obtained at the EWPT if it is of the
first order and proceeds via bubble nucleation. A reliable evaluation of the
order parameter of the EWPT, {\it i.e.} of the vacuum expectation values
(VEV's) of the Higgs fields,
requires a detailed calculation which has
been carried out by now only for the Standard Model (SM) \cite{zwirner}.
Regarding the  extensions of the SM, in particular the Minimal Supersymmetric
Standard Model (MSSM) which we consider in the present paper, no definite
answer is at hand.

As far as B-violating interactions are concerned, it is well-known
that the baryon number is badly violated by anomalous interactions at
temperatures above or at the electroweak scale \cite{kuz}. Moreover, the
requirement that anomalous interactions freeze out just after the
completion of the EWPT, otherwise they may wipe out any baryon asymmetry
previously created, translates into a very strong bound on the
VEV's of the Higgs fields
\beq
\frac{v({\rm T_C})}{{\rm T_C}} \simgreat 1
\eeq
where ${\rm T_C}$ is the critical temperature.
The interesting fact
is that this limit provides a constraint on  the Higgs mass spectrum
of the theory.
For instance, it seems by now established that the LEP lower bound $m_{h}
>60$ GeV \cite{lep} on the mass of the Higgs scalar is
incompatible with the weak baryogenesis scenario in the SM. In the MSSM
the LEP lower bound on the lightest Higgs scalar is presently around 43
GeV. The analysis which have been carried out by now \cite{zz,nan}
in the context of the one-loop approximation for the finite temperature
effective potential indicate that the condition (1) is indeed a strong
constraint even in the MSSM, but it is not clear
whether this limit rules out or not
the possibility of generating the baryon asymmetry of the Universe at
the EWPT \cite{zz}.
However it is likely that the actual upper bound  on the lightest Higgs
mass coming from (1) is not very far from the experimental lower one.
Moreover, if one considers also the dependence of $v({\rm T_C})/{\rm
T_C}$ on the stop-sbottom masses and the $\rho$ parameter, then a small
region of parameter space is left, which will be explored by LEP
II \cite{zzwi}.

The third of Sakharov's conditions is the presence of CP violation,
which is necessary in order that B-violating interactions do not produce
the same amount of baryons and of antibaryons. In the SM it seems that
the amount of CP violation is too small to give rise to the observed
asymmetry \cite{shapcp}.
Simple extensions of the SM with an enlarged Higgs sector,
 like the two Higgs model or the MSSM, contain
new sources of CP violation and for such a reason can provide nice
scenarios for baryogenesis at the weak scale. In particular, the
MSSM contains two extra CP violating phases with respect to the SM.
The requirement that these phases provide the necessary amount of CP
violation for the generation of the BAU, gives rise to additional
strong constraints on the parameter space of the model \cite{cohen}.
Indeed, the electric dipole moment (EDM) of the neutron must be larger
than $10^{-27}$ e-cm, while  an
improvement of the current experimental bound on it by one order of
magnitude would constrain the
lightest chargino and the lightest neutralino to be lighter than 88 and
44 GeV, respectively \cite{cohen}.

In the MSSM another source of CP violation may emerge at finite
temperature. As was shown in ref. \cite{cp}, one-loop effects at a
temperature T = ${\cal O}$(100) GeV may induce a non zero relative phase,
$\delta$,
between the VEV's of the two Higgs that is, these effects may induce
spontaneous CP violation (SCPV). The phase $\delta$ varies with the
VEV's and with the temperature and, as T goes to zero, it becomes
trivial ($\delta=0,\:\pm\pi$). Due to this fact,
the `spontaneous' phase $\delta$ can assume all the values in the range
$0\leq \delta\leq \pi$
and, contrary to the `explicit' phases considered in ref. \cite{cohen},
does not receive any bound from current experimental constraints
on the neutron EDM.

In this paper we will consider the implications of the finite temperature
SCPV on baryogenesis. The spontaneous breakdown of CP can lead to different
cosmological
scenarios according to the nature of the EWPT. After briefly discussing
the case in which the EWPT is of the second order and the related
problem of the formation of domain walls, we investigate the relevance of
the new source of CP violation
for the so called spontaneous baryogenesis scenario
\cite{nel,abel} when the EWPT is of
the first order and proceeds via bubble nucleation.
The presence of a space-time varying phase $\delta$, can induce
a shift in the energy levels between baryons and antibaryons and bias
the anomalous, B violating, interactions, giving rise to a
 non vanishing baryon asymmetry. The presence of expanding bubbles in the
thermal bath is essential to this mechanism, since the baryon asymmetry is
generated inside the bubble walls where the phase $\delta$ changes its value
and the B violating processes are active.

Since the phase $\delta$ can take values of order one, the baryon
number creation mechanism by a single bubble turns out to be very
efficient. For comparison, the `explicit' phases considered in
\cite{cohen,nel} are constrained to be of order $10^{-2}-10^{-3}$ by the
bounds on the EDMN, and the baryon  number generated by a single bubble
is estimated to be $10^{-2}-10^{-3}$ times the one generated if SCPV is
present. However, if CP is broken {\it only} spontaneously, then the
phase $\delta$ can take two opposite values, corresponding to two {\it
exactly} degenerate vacua.
This degeneracy between the two vacua in $\pm \delta$
would induce an equal number of nucleated bubbles carrying phases with
opposite signs, which in turn generate baryon asymmetries of opposite
signs.
The BAU obtained averaging over the entire volume of the
Universe would then be zero. We find that the introduction of very
small explicit phases, of order  $10^{-5}-10^{-6}$, lifts the
degeneracy, leading to a difference between the nucleation rates
of the two kinds of
bubbles, and possibly to a baryon asymmetry of the right order of magnitude.
We wish to stress that the `spontaneous' phase $\delta$ and the
`explicit' ones play very different roles in this scenario, the former
being the real source of CP violation necessary to the primary
production  of
baryons, and the latter lifting the degeneracy between otherwise
equivalent vacua in order to achieve a global bias of the primary production.

{}From the phenomenological point of view, such very small phases
give rise to negligible contributions to the  neutron EDM.
 Moreover, they may be generated through the Renormalization
Group Equations (RGE's) due to the complexity of the Yukawa couplings
\cite{dug} even if the soft supersymmetry (SUSY) breaking parameters and the
$\mu$ parameter of the superpotential, are chosen to be real at the
SUSY breaking scale. In this picture, the only sources of CP
violation in the MSSM turn out to be the phase in the
Cabibbo-Kobayashi-Maskawa matrix and the strong parameter $\theta$.

The paper is organized as follows. In Section {\bf 2} we discuss how finite
temperature effects can induce SCPV in the Higgs sector. In Section
{\bf 3} we analyze the implications for baryogenesis of the SCPV
in the cases in which the EWPT is of the first or the second order.
Finally, in Section {\bf 4} we present our Conclusions.
\vspace{1.cm}
\\

{\large {\bf 2. Spontaneous {\cal CP}-Violation in the MSSM at Finite
Temperature}}

\vskip 0.8 cm

In this Section we briefly review the recent results on
 the spontaneous {\cal CP}-violation in the scalar
sector of the MSSM as a finite temperature effect \cite{cp}.

As already pointed out in refs. \cite{pom}, radiative corrections may
induce SCPV in the MSSM even at zero temperature. However it comes out that
the  corresponding region of the parameter space is not physical.
The point is that spontaneous {\cal CP}-violation can be implemented
radiatively only if a pseudoscalar with zero
tree-level mass exists, as it was shown on general grounds by Georgi
and Pais \cite{Georgi}. In the MSSM this implies the existence of a very
light Higgs (with a mass of a few GeV, given by one--loop contributions)
\cite{pom}, which has been excluded by LEP data
\cite{lep}.

As was discussed in ref. \cite{cp}, one-loop effective potential at  finite
temperature may allow SCPV at T=${\cal O}$(100) GeV and satisfy the
experimental
bounds on the mass of the Higgs pseudoscalar at the same time.

The most general gauge invariant scalar potential for the
two--doublets model, along the neutral components, is given by
\begin{eqnarray}
V &=& {m_{1}}^{2}|H_1|^2 + {m_{2}}^{2}|H_2|^2 - ({m_{3}}^{2} H_1 H_2 +
{\rm h.c.})
+\lambda_1 |H_1|^4 +\lambda_2 |H_2|^4  \nonumber\\
&+& \lambda_3 |H_1|^2 |H_2|^2 + \lambda_4|H_1 H_2|^2
+\left[\lambda_5 (H_1 H_2)^2  + \lambda_6 |H_1|^2 H_1 H_2 +\lambda_7 |H_2|^2
H_1 H_2 + {\rm h.c.}\right].
\end{eqnarray}
The tree level scalar potential for the MSSM can be obtained from eq.
(2) by imposing the relations
\begin{eqnarray}
\lambda_1 &=& \lambda_2\:\:=\:\:\frac{1}{8}({g_2}^{2}+{g_1}^{2}),\nonumber\\
\lambda_3 &=& \frac{1}{4}({g_2}^{2}-{g_1}^{2}),\nonumber\\
\lambda_4 &=& -\frac{1}{2}{g_2}^2, \nonumber \\
\lambda_5 &=& \lambda_6\:\:=\:\:\lambda_7\:\:=\:\:0,
\end{eqnarray}
where $g_2$ and $g_1$ are the gauge couplings of $SU(2)_L$ and $U(1)_Y$
respectively, and $m_3^2$ may always be assumed to be real
\cite{dug}. The potential in eq. (2) can be minimized by complex
vacuum expectation values (VEV's)
\begin{equation}
\langle H_1\rangle=v_1,  \:\:\:
\langle H_2\rangle=v_2 \: {\rm e}^{i\delta}, \:\:\: \delta\neq 0, \: \pi
\end{equation}
if, and only if
\begin{equation}
\lambda_5>0\:, \:\:\:\:\: {\rm and }\: \:
-1 < \cos \delta = \frac{m_3^2 - \lambda_6 v_1^2 -\lambda_7 v_2^2}
{4 \lambda_5 v_1 v_2} < 1.
\end{equation}
{}From eq. (3) and (5) we immediately realize that
no spontaneous CP violation
can occur in the scalar sector of the MSSM at the tree-level
since $\lambda_5=0$ .

 The one-loop contribution at finite temperature to the effective
potential can be decomposed into the sum of a T=0 and a ${\rm T}\neq0$ term:
\begin{eqnarray}
\Delta V_{{\rm T}=0} &=& \frac{1}{64 \pi^2} {\rm Str} \left\{ {\cal M}^4
\left(\ln\frac{{\cal M}^2}{{\cal Q}^2}-\frac{3}{2}\right)\right\},\\
\Delta V_{{\rm T}\neq 0} &=& \Delta V^{bos.}_{{\rm T}\neq 0} + \Delta
V^{ferm.}_{{\rm T}\neq 0}.
\end{eqnarray}
Defining $a^2_{b,(f)}\equiv{\cal M}_{b,(f)}^2/{\rm T}^2$, where
${\cal M}_{b,(f)}$ is the bosonic (fermionic) mass matrix, the ${\rm T}\neq 0$
contributions may be written  as

\begin {eqnarray}
\Delta V_{{\rm T} \neq 0}^{bos.} &=& {\rm T}^4\:\: {\rm Tr}' \left[
\frac{1}{24} a_b^2-
\frac {1}{12 \pi^2}(a_b^2)^{3/2}  - \frac{1}{64 \pi^2}a_b^4
\ln\frac{a_b^2}{A_b} \right. \nonumber \\
&-& \left. \pi^{3/2}\:\: {\sum}_{l=1}^{\infty}
(-1)^l \:\:\frac{\zeta(2l+1)}{(l+1)!}\:\:
\Gamma\left(l+\frac{1}{2}\right)\:\:
\left(\frac{a_b^2}{4 \pi^2}\right)^{l+2} \right],
\end{eqnarray}
\begin{flushright}
($a_b < 2 \pi$)
\end{flushright}
\begin{eqnarray}
\Delta V_{{\rm T} \neq 0}^{ferm.}&=& {\rm T}^4\:\: {\rm Tr}' \left[
\frac{1}{48}a_f^2 +
\frac{1}{64 \pi^2}a_f^4 \ln\frac{a_f^2}{A_f}
\right.
\nonumber\\
&+& \left. \frac{\pi^{3/2}}{8} \:\:
\sum_{l=1}^{\infty} (-1)^l\:\: \frac{1-2^{-2l-1}}{(l+1)!}\:\:
\zeta(2l+1)
\:\:\Gamma\left(l+\frac{1}{2}\right) \left(\frac{a_f^2}{\pi^2}\right)^{l+2}
\right],
\end{eqnarray}
\begin{flushright}
($a_f<\pi$)
\end{flushright}
\vskip 0.2 cm
where $A_b = 16 A_f = 16 \pi^2{\rm exp}(3/2 - 2 \gamma_E)$, $\gamma_E =
0.5772$, $\zeta$ is the Riemann function, and ${\rm Tr}'$ properly
counts the degrees of freedom.
Eqs. (8) and (9) give an exact representation of the complete one--loop
effective potential at finite temperature \cite{Dolan} for $a_b< 2\pi$
and $a_f<\pi$, respectively.

As long as SUSY is exact, $\lambda_{5, 6, 7}$ are zero at any
order in perturbation theory. So the only renormalization to these
parameters may come from the soft SUSY breaking sector. In particular,
we find that the dominant contributions are those coming from the
gaugino mass terms, $M_{1, 2}$, which enter the charginos and
neutralinos mass matrices, and by the sfermion mass terms
$\tilde{m}_{t_{L, R}}$ which appear in the stop mass matrices. In the
following we will evaluate the one-loop renormalizations to
$\lambda_{5, 6, 7}$ and $m_3^2$ at finite temperature, including in the
effective potential (7) the mass matrices of charginos,
neutralinos and stops.

Taking $\tilde{m}_Q^2=\tilde{m}_U^2$ the stop mass matrix takes the
convenient form
\begin{equation}
a_t^2=a_Q^2\cdot {\bf 1}  + \tilde{a}_t^2
\end{equation}
where $a_Q^2\equiv \tilde{m}_Q^2/{\rm T}^2$, ${\bf 1}$ is the identity matrix,
and $\tilde{a}_t^2$ is the field--dependent part of the mass matrix, {\it
i.e.} $\tilde{a}_t^2\rightarrow 0$ as the fields vanish.
Summing up all the terms in eq. (8) in $\tilde{a}_t^4$, $\tilde{a}_t^6$
and  $\tilde{a}_t^8$,
the following contributions are obtained from the stop \cite{cp}
\beqra
\Delta {m_3^{(s)}}^2 &=& + 3 h_t^2 A_t {\rm T} a_\mu \left[ \frac{1}{8 \pi a_Q}
+ \frac{1}{16 \pi^2}
\left(\ln \frac{{\cal Q}^2}{A_b {\rm T}^2} + \frac{3}{2}\right) +
4 {\cal B}_4 [a_Q^2] \right],\\
 & &\nonumber \\
\Delta \lambda_5^{(s)}&=&-12 h_t^4 \frac{A_t^2 a_\mu^2}{{\rm T}^2}
\left[ {\cal B}_8[a_Q^2] + \frac{1}{256 \pi a_Q^5} \right],\\
& & \nonumber \\
\Delta \lambda_6^{(s)}&=& - 6 h_t^2 \frac{A_t a_\mu}{{\rm T}}
\left[ \frac{3}{4} (g_2^2+g_1^2)\left({\cal B}_6[a_Q^2]-\frac{1}{192 \pi
a_Q^3}\right)\right. \nonumber \\
&+&\left. 4 h_t^2 a_\mu^2 \left( {\cal B}_8[a_Q^2] +
\frac{1}{512 \pi a_Q^5}\right)\right],\\
& & \nonumber \\
\Delta \lambda_7^{(s)}&=& - 6 h_t^2 \frac{A_t a_\mu}{{\rm T}}
\left[ \left(6 h_t^2 -\frac{3}{4}(g_2^2+g_1^2)\right)
\left({\cal B}_6[a_Q^2]-\frac{1}{192 \pi a_Q^3} \right)\right. \nonumber
\\
&+&\left. 4 h_t^2 \frac{A_t^2}{{\rm T}^2} \left( {\cal B}_8[a_Q^2] +
\frac{1}{512 \pi
a_Q^5}\right)\right],
\eeqra
where we have defined $a_\mu\equiv \mu/{\rm T}$, $\mu$ is the coefficient of
the
term $H_1H_2$ in the superpotential, and
\begin{equation}
{\cal B}_{2n}[a_Q^2]\equiv \pi^{3/2} \sum_{l={\rm max}[1,n-2]}^{\infty}\:\:
(-1)^l\:\:
\frac{\zeta(2l+1)}{(l+2)!}\:\:\Gamma\left(l+\frac{1}{2}\right) {l+2 \choose n}
\frac{(a_Q^2)^{l+2-n}}{(4 \pi^2)^{l+2}}
\end{equation}
\begin{flushright}
$n=2, 3, 4\ldots\:\:\:\:\:\:\:\:\:a_Q<2 \pi$.
\end{flushright}
The series in (15), having terms of alternate signs, may be easily
evaluated numerically.
With the choice, $|\mu^2|=|M_1^2|=|M_2^2|=a_\mu^2 {\rm T}^2$,
the chargino and neutralino
contribution are obtained in a completely analogous way and read
\beqra
\Delta {m_3^{(c)}}^2&=& + \:\:  g_2^2 \: M \: \mu
\left[ \frac{1}{8 \pi^2}
\left(\ln \frac{{\cal Q}^2}{A_f {\rm T}^2} +\frac{3}{2}\right) +
8 {\cal F}_4[a_{\mu}^2] \right]\\
\Delta \lambda_5^{(c)}&=& 8 \:g_2^4
\left(\frac{ M \: \mu}{{\rm T}^2}\right)^2 {\cal F}_8[a_{\mu}^2],\\
\Delta \lambda_6^{(c)}&=&\Delta \lambda_7^{(c)}\:\:=
- \:\: 4\:\: g_2^4 \:\: \left(\frac{M \:\mu}
{{\rm T}^2}\right)^2 \left[ 3 {\cal F}_6[a_\mu^2] +
4 a_\mu^2 {\cal F}_8[a_\mu^2]\right],
\eeqra
and
\beq
\Delta {m_3^{(n)}}^2 = \frac{(g_2^2 +g_1^2)}{2 g_2^2} \Delta
{m_3^{(c)}}^2,\:\:\:\:\:\:\:\:\:\:\:
\Delta \lambda_i^{(n)} = \frac{(g_2^2 +g_1^2)}{2 g_2^2}\Delta
\lambda_i^{(c)},\:\:\:\:\:i=5,6,7.
\eeq
where
\beq
{\cal F}_{2n}[a_{\mu}^2]=\frac{\pi^{3/2}}{8}
\sum_{l={\rm max}[1,n-2]}^\infty\:\: (-1)^l\:\: \frac{1-2^{-2l-1}}{(l+2)!}
\:\:\zeta(2l+1)\:\: \Gamma\left(l+\frac{1}{2}\right)
{l+2 \choose n} \:\:\frac{(|a_{\mu}|^2)^{l+2-n}}{(\pi^2)^{l+2}}
\eeq
In ref. \cite{cp}, it is shown that the conditions for SCPV at finite
temperature, {\it i.e.}
\beq
\Delta \lambda_5 >0,
\eeq
and
\beq
\left| \cos \delta = \frac{\bar{m}_3^2 -\Delta \lambda_6 v_1^2({\rm T}) +
\Delta \lambda_7 v_2^2({\rm T})}{4 \Delta \lambda_5 v_1^2({\rm T}) v_2^2(
{\rm T})}\right|<1
\eeq
where $\bar{m}_3^2 \equiv m_3^2 +\Delta m_3^{2(s)}+\Delta m_3^{2(n)}+
\Delta m_3^{2(c)}$ and $\Delta \lambda_i\equiv
\Delta \lambda_i^{(s)}+\Delta
\lambda_i^{(c)}+\Delta \lambda_i^{(n)}$ ($i= 5, 6, 7$)
can be satisfied, for temperatures around the electroweak scale,
in a wide region of the parameter space compatible with the
present experimental bound on the mass of the Higgs pseudoscalar.
Typical numerical values for $\Delta \lambda_5$ are around $10^{-5}$,
whereas $\Delta \lambda_6, \Delta \lambda_7 \approx 10^{-4}$.
This means that in order to have spontaneous CP violation, $\bar{m}_3^2$
 must be of order
$\Delta \lambda_6 \: v_1^2({\rm T}) +\Delta \lambda_7 \: v_2^2({\rm T})$
within a
10 percent, see eq. (22).

This spontaneous CP breaking can lead to different cosmological implications
according to the nature of the EWPT.
If the latter is of
the second order, all the quantities in eq. (22) will evolve smoothly with
the temperature, so that there will be a time interval during which the
conditions
(21) and (22) are satisfied, and $\cos \delta$ varies from 1 to --1.
As the temperature decreases below a critical value ${\rm T}_{rest}$,  the
${\rm T}$-dependent quantities in eq. (22) become such that $|\cos \delta|>1$
and CP is restored.

On the other hand, if the EWPT is of the first order and proceeds
by bubble nucleation, the temperature keeps constant  until  all the
Universe is in the broken phase,
then  $\bar{m}^2_3$ and $\Delta \lambda_i$  are fixed during the phase
transition, whereas $v_1 $ and $v_2$
 change their values from zero
to $v_1({\rm T}_c)$ and $v_2({\rm T}_c)$ when a bubble wall passes through
a fixed point.
Any steady observer will then experience  a change in the phase $\delta$ from
0 to $\pi (-\pi)$, if, as we assume, inside the bubble the modulus of
the ratio in eq. (22) is greater than one.
In the following Section we investigate the cosmological implications of
the two different scenarios outlined above.
\vspace{1cm}\\

{\large{\bf 3. Spontaneous CP violation and baryogenesis}}

\vskip 0.8 cm

As we have discussed in the previous Section, the evolution of the
effective potential as the temperature decreases may induce a
non vanishing relative phase between the two Higgs VEV's. Such a new source
of CP violation may be relevant for the generation of the baryon
asymmetry of the Universe during the EWPT by means of the so called
`spontaneous baryogenesis scenario' \cite{nel,abel}. The main underlying
idea of this mechanism is that a space-time varying phase $\delta$ may
induce an interaction of the form
\beq
{\cal L}_{int}\sim \partial_\mu \delta \:\:J^\mu.
\eeq
If the current $J^\mu$ is not orthogonal to the baryonic one, then the
interaction term (23) shifts the energy levels of baryons relative to
antibaryons. The time derivative of the phase $\delta$ acts as an
effective chemical potential (usually called `charge potential') for the
baryon number. If an interaction like that in eq. (23) is active when
also baryon number violating interaction are in equilibrium, then the
thermodynamic evolution of the system leads to a non zero baryon
asymmetry.

In the case of baryogenesis at the electroweak scale, the
necessary B violation is provided by the non perturbative
`sphaleron-like' processes \cite{klink,kuz}, which violate the fermion
number, B+L, and conserve the orthogonal combination, B-L (L is
the lepton number), since it  is
anomaly free in the SM as well as in the MSSM.
An interaction of the type (23) emerges in the MSSM if one chooses the
fermion basis in such a way that at any time and at any point in space
the fermion masses are real. This implies that a space-time dependent
rotation has to be done on the fermion fields in order to remove the phase
$\delta$ from the Yukawa couplings, at the price of introducing new
interactions of the form (23) coming from the fermionic kinetic terms
\cite{nel}. In general the rotation of the fermion fields will be
anomalous, so it will induce extra terms of the Chern-Simons type in
the action. These terms may bias the sphaleron transition, giving an
additional contribution to the generation of the baryon asymmetry
\cite{turok}. However, as was discussed on ref. \cite{abel} this effect
gives rise to a baryon asymmetry which is a few percent of the one
generated by the interaction (23), so we will ignore the Chern-Simons
contributions in the following.

Following ref. \cite{nel}, we can calculate the density of a species
{\it i} under the assumption that all the particles are in thermal
equilibrium
\beq
\rho_i = k_i \left( \dot{\theta_i} + ({\rm B}_i - {\rm L}_i) \mu_{\rm B-L}
+ {\rm Q}_i \mu_{\rm Q} \right) \frac{{\rm T}^2}{6},
\eeq
where $k_i$ counts colour and spin degrees of freedom, and factor two
for bosons with respect to fermions, $\theta_i$ is the phase by which
the fermion species has been rotated, $\mu_{\rm B-L}$ and $\mu_{\rm Q}$ are
the chemical potentials of the conserved quantum numbers B--L and Q,
respectively. If we define $H_{1,2}({\rm T})=v_{1,2}({\rm T}) {\rm exp}(i
\theta_{1,2}({\rm T}))$, we see that a change in the phase
$\delta =\theta_1 +\theta_2$ will be shared between $\theta_1$ and
$\theta_2$ according to the variation
\beq
{\rm d}\theta_1=\frac{v_2^2 ({\rm T})}{v_1^2({\rm T}) + v_2^2({\rm T})}
{\rm d}\delta ,\:\:\:\:\:\:
{\rm d}\theta_2=\frac{v_1^2 ({\rm T})}{v_1^2({\rm T}) + v_2^2({\rm T})}
{\rm d}\delta
\eeq
which is orthogonal to the unphysical Goldstone mode.
Imposing the constraints $\rho_{\rm B-L}=\rho_{\rm Q}=0$,
and assuming that the sphaleron-like transitions are in equilibrium too,
one can solve the system (24) for the lepton and baryon number densities
$\rho_{\rm L}$ and $\rho_{\rm B}$ \cite{nel}
\beq
\rho_{\rm L}=\rho_{\rm B}=-12\frac{6+{\rm n}}{111 + 13 {\rm n}}
\frac{{\rm T}^2}{6} \dot{\delta}
\eeq
where n is the number of light charged scalars.
However during the EWPT the simplifying assumptions leading to eq. (26)
have to be revisited. In particular, the only right-handed fermion which
might be in thermal equilibrium is the top. Moreover, since
 the typical time scale of the CP violation in this system
is ($10^{-2} - 1) {\rm GeV}^{-1}$, the rate for anomalous baryon number
violation is too slow to allow baryon number to reach its equilibrium
density (24). For such reasons it seems more appropriate to
make use   \cite{nel} of the relaxation equation \cite{sakita}
\beq
\dot{\rho}_{\rm B} = - 9 \Gamma_{\rm B} \frac{\mu_{\rm B}}{{\rm T}},
\eeq
where
\beq
\Gamma_{\rm B} = \kappa \alpha_{\rm W}^4 {\rm T}^4 {\rm e}^{-4 \pi v({\rm
T})/g_{{\rm W}} {\rm T}}
\eeq
is the rate for the baryon number violation per unit volume,
$v({\rm T})=(v_1^2({\rm T}) + v_2^2({\rm T}))^{1/2}$, the prefactor
$\kappa$
is evaluated
to be in the range 0.1--1 by numerical simulations \cite{amb} and
$\mu_{\rm B}$ is the charge potential for the baryon number. Setting the
net densities of the light, right-handed particles to zero, $\mu_{\rm
B}$ reads \cite{nel}
\beq
\mu_{\rm B}=-\frac{4}{3}\frac{6 + {\rm n}}{25 + 4 {\rm n}} \dot{\delta}.
\eeq

Eq. (27) will be integrated in Section {\bf 3.2} when we will estimate the
baryon asymmetry produced under the hypothesis of a first order
EWPT. Before turning to that scenario, which is the more promising from
the point of view of baryogenesis, we want to make some comments on the
cosmological implications of the spontaneous CP violation in the MSSM
when the phase transition is of the second order.
\vspace{1. cm}
\\
{\large {\bf 3.1 Second order EWPT and domain walls}}

\vskip 0.8 cm

If we assume that the EWPT is of the second order, as expected for large
values of the mass of the lightest Higgs, the spontaneous CP violation
will lead to the formation of domain walls separating regions with
opposite signs of the  phase $\delta$, whose cosine is given by eq.
(22) \cite{okun}. Domain walls are an unavoidable prediction of models
in which global discrete symmetries are spontaneously broken \cite{vil}.
Since their  energy density grows faster than that of the radiation, at
a certain temperature they begin to dominate the evolution of the
Universe, destroying its homogeneity and the isotropy of the relic
radiation. However, if the discrete symmetry is broken only in a
limited temperature interval, as for example in the scenario proposed in
ref. \cite{kuzdw}, then the domain walls will decay as the symmetry is
restored. If the temperature of symmetry restoration is higher than that
at which the walls begin to dominate over radiation any dangerous
consequence for the evolution of the Universe is avoided.
This is precisely what happens in the present case, since CP is restored at a
temperature that is of the
order of $1-\lambda_5/\lambda_6$ times the temperature at which it is
broken, {\it i.e.} ${\rm T}_{rest} = {\cal O}(100)\:{\rm GeV}$, whereas the
temperature at which the walls begin to dominate is  about $10^{-4}
(v/100\: {\rm GeV})^{3/2}\: {\rm MeV}$.

Domain walls may also play a  role in baryogenesis, as
for instance in the left-right symmetric model considered in refs.
\cite{hol,henry}. In principle, also in the model we are considering one could
think to investigate on  the possibility of implementing the spontaneous
baryogenesis mechanism. Indeed, the phase changes from $\delta$ on one
side of the wall to $-\delta$ on the other side, so that there is a
net change of $2 \delta$ at any point when the wall passes by it. The
motion of the wall can be obtained if the degeneracy between the two
vacua is slightly lifted, for example by introducing small `explicit'
phases $\phi$. At the temperature at which the difference in energy
density between the two vacua  equals the energy density of the wall
itself, which is of the order of $(\lambda_5 \phi M_P^{1/2}
v^{5/2})^{1/3}\sim 100\: {\rm GeV}$ for $\phi \sim 10^{-3}$,  the walls
start moving in the direction of the energetically disfavoured domain.
A necessary condition to make the
spontaneous baryogenesis scenario work is however that the
baryon number violating processes are active inside the walls and
suppressed outside, otherwise the baryon asymmetry would be wiped out.
Recalling the condition in eq. (1), this  could be achieved if for some
temperature interval the ratio $v/{\rm T}$ were sensibly less than
unity inside
the wall and greater outside. Unfortunately this is not the case in this
model, since the difference between the two VEV's is of the order of
$(\lambda_6/g^2)^{1/2} v\sim 10^{-2} v$, consequently the B violating
processes freeze out inside the wall and outside it contemporary.
\vspace{1 cm}

{\large \bf 3.2 First Order EWPT and Spontaneous Baryogenesis}

\vskip 0.8 cm

We now turn to the discussion of the consequences of the SCPV in the case in
which the EWPT is of first order and proceeds via bubble nucleation.
In our scenario the spontaneous baryogenesis mechanism is implemented
 through the change of the VEV's phase $\delta$
inside the bubble walls where the VEV's moduli change from 0 to
$v_{1,2}({\rm T_C})$.

If one approximates the exponential of eq. (28) by a step
function and takes into account
that the temperature during the EWPT keeps constant, one can easily
integrate eq. (27) and find the baryon number density to
entropy ratio \cite{nel} induced by a {\it single} bubble
\begin{equation}
B\equiv \left(\frac{n_B}{s}\right)_{local} \simeq 3\times  10^{-7}\: \left(
\frac{\Delta \delta}{\pi}\right).
\end{equation}
The local baryon number density to entropy
 ratio (30) generated by the passage of the wall would
be erased if the anomalous interactions were still active in the true vacuum.
If in the middle of the bubble $v({\rm T})/{\rm T} \simgreat 1$,
the sphaleron rate
(28)
is safely suppressed.
This bound translates into an upper limit on the mass of the lightest higgs
 scalar in the MSSM. As we have discussed in the Introduction, it is not
clear whether this limit is or not compatible with LEP data
\cite{lep}.
In order to get a definitive answer, a detailed calculation along the
lines of ref. \cite{zwirner}
would be necessary.
In the following we assume that the first order phase transition is
sufficiently strong so that the bound (1) is satisfied.
Moreover, we will define the bubble wall as the
region in which sphalerons are active, that is, in which $0<v({\rm
T})/{\rm T} \simless 1$.

Since  the temperature  keeps constant throughout the
phase transition, the change
of the phase $\delta$ at a given point is induced by the change of the
VEV's $v_{1,2}({\rm T})$ as the bubble wall passes  through that point.
More precisely, one can easily infer from eq. (22) that,
as the VEV's change from $v_{min}$ to $v_{max}$, where
\beq
v_{min}^2=\frac{\bar{m}_3^2}{\cos^2 \beta}\:\: \frac{1}{\lambda_6 +
\lambda_7 \tan^2 \beta -4 \lambda_5 \tan \beta}\:\:,
\eeq
$$
v_{max}^2=\frac{\bar{m}_3^2}{\cos^2 \beta}\:\: \frac{1}{\lambda_6 +
\lambda_7 \tan^2 \beta +4 \lambda_5 \tan \beta}
$$
the relative phase $\delta$ changes from $0$ to $\pm \pi$ (we are
assuming here that $\tan \beta$ keeps constant at any temperature) .
If $v_{max}/{\rm T}\simless 1$, the variation of $\delta$ lies entirely inside
 the bubble wall, so that
$\Delta \delta$ in eq. (30) may be as large as $\pi$.
In ref. \cite{cp} it is shown that the requirement $v_{max}/{\rm T}\simless 1$
is fulfilled in a phenomenologically acceptable region of the
parameter space. On the other hand, from eq. (31) we read that this
condition implies $0<\bar{m}_3^2 \simless \lambda_{6, 7} {\rm T}^2$, which
in turn induces a strong correlation between the parameters $m_A^2$ and
$\tan \beta$ \cite{cp}. In fig. (1) we see that this requirement fixes
the pseudoscalar mass with an accuracy of roughly 10 percent, which
gives a measure of the `fine tuning' needed by this mechanism.
In fig. (2) we have allowed the other free parameters of the model to
vary in the
range allowed by the condition (21) \cite{cp}, and we have obtained the
absolute upper and lower bounds to the pseudoscalar mass. The dashed
line represents the experimental lower bound \cite{lep}, so that we can
also see
that values of $\tan\beta$ less than $\sim$7 are excluded.

Note that in the case of explicit CPV like that considered in ref. \cite{nel}
we have
$\Delta \delta ={\cal O}(1)\delta_{CP} $ where $\delta_{CP} $ is bounded to
 values less than $10^{-2}-10^{-3}$ by the experimental limits on the
electric dipole moment (EDM) of the neutron, which do not apply to
the present case.
So we can read from eq. (30) that the baryon production by the {\it
single}  bubble
with the mechanism discussed in this paper is at least two or three orders of
magnitude more efficient than in the case of an explicit CPV.

As we have discussed in Section  {\bf 3.1}, for any set of values of the
temperature and the other parameters of the model, the effective
potential has a double degeneracy in $\pm \delta$  as long as eq. (22) is
satisfied and there are no new complex phases other than the one in the
Cabibbo-Kobayashi-Maskawa matrix. Since we are assuming that the Higg's
VEV inside the bubble is greater than $v_{max}$, {\it i.e.} CP is
restored, the vacuum in the broken phase is now unique, unlike the
situation discussed in Section {\bf 3.1}. On the other hand,
inside the bubble walls, where
the VEV is between $v_{min}$ and $v_{max}$, the condition (22) is
satisfied, and then the ambiguity between the two signs is present.
Stated in other words, the phase $\delta$ may follow two different paths
as the VEV's change inside the walls, namely, from 0 to $\pi$ or from 0
to $-\pi$. This results in two possible signs for $\dot{\delta}$, and
then to the nucleation of two different kinds of bubbles, say `plus' and
`minus' bubbles which, according to eq. (30), create baryons of opposite
signs. Since there is no way to prefer one kind of bubble relative to
the other one, the net baryon asymmetry, averaged over the entire Universe,
will be zero.

In the following Subsection we show that allowing the soft SUSY breaking
parameters and the $\mu$ parameter to take complex values, the effective
potential takes on a $\sin\delta$ dependence, and then one sign of
$\delta$ becomes energetically favourite with respect to the opposite
one. In this way the two kinds of bubbles have slightly different
surface tensions, and then free energies, so that their nucleation rates
are no more equal. As a result, an abundance of one kind of bubbles
relative to the  other, and then a non zero average baryon asymmetry, is
achieved. We will also find that the job may be done by  `explicit'
phases as little as $10^{-5}-10^{-6}$, which give a
negligible contribution to the EDMN.
\vspace{1.cm}

{\bf 3.2.1 Choosing the good bubbles}

\vskip 0.8 cm

Defining
\beq
A=|A| {\rm e}^{i\phi_A},\:\: \:\: B=|B| {\rm e}^{i \phi_B}
\eeq
where $A$ and $B$ are the
trilinear and bilinear soft SUSY breaking parameters is it always possible
to rotate the phase of the parameter
$\mu$ taking
\beq
\mu=|\mu| e^{-i \phi_B},
\eeq
so that $m_3^2=\mu B $ is real.
$A$ and $\mu$ enter the one-loop corrections for $\bar{m}_3^2$
 and $\lambda_{5,6,7}$ (see Section  {\bf 2}),
so these parameters take phases which, if $\phi_A, \phi_B \ll 1$, are given by
\begin{eqnarray}
\phi_3&=&\frac{|\Delta m_3^{(s)}|}{|\bar{m}_3^2|}\phi_A -
\frac{|\Delta m_3^{(s)}|+|\Delta m_3^{(n)}|+|\Delta m_3^{(c)}|}
{|\bar{m}_3^2|}\phi_B,\nonumber\\
\phi_i&=&\frac{|\Delta \lambda_i^{(s)}|}{|\lambda_i|} \phi_A -\phi_B,
\:\:\:\:\:\: i=5,6,7,
\end{eqnarray}
where we have defined $\bar{m}_3^2=|\bar{m}_3^2| {\rm e}^{-i \phi_3},
\:\:\lambda_5=|\lambda_5| {\rm e}^{-2i \phi_5}$ and $\lambda_{6,7}
=|\lambda_{6,7}|
{\rm e}^{-i \phi_{6,7}}$, and approximated $\sin \phi_{A, B}\simeq \phi_{A,
B}$.
The scalar potential now has also terms depending on $\sin \delta$,
which are given by
\begin{eqnarray}
\Delta V (\sin \delta) &\simeq&
2 \sin  \delta \:\:v_1({\rm T})\:\: v_2({\rm T})\times\nonumber\\
&&\left[|\bar{m}_3^2| \:\:(\phi_5-\phi_3) -|\lambda_6| \:\: v_1^2({\rm
T})
\:\: (\phi_5-\phi_6)\right.\nonumber\\
&-&\left.|\lambda_7| \:\: v_2^2({\rm T})\:\:(\phi_5-\phi_7)\right].
\end{eqnarray}
For fixed $v_{1, 2} ({\rm T})$ there are two minima in $\delta$. The potential
(35) slightly removes the degeneracy between them, that is, it
distinguishes between `plus' and `minus' bubbles.

We now come to an estimation of the
abundance of one type of nucleated bubbles
relative to the other one, due to the breaking of
the degeneracy.
Inside a comoving volume coincident with the horizon volume , the ratio between
the number of `plus' and `minus' bubbles is
\beq
\frac{N_{+}}{N_{-}}={\rm e}^{-\Delta F/{\rm T}},
\eeq
where $\Delta F$ is the difference between the free energies of the two
kinds of critical bubbles.
The free energy of a bubble of radius $R$ is given  by the sum of the
volume and surface energies, $F=4 \pi R^3 V /3 - 4 \pi R^2 \sigma$,
where $V$ is the
density energy of the true vacuum and $\sigma$ the surface energy density.
Reminding that the radius of a critical bubble, {\it i.e.} the minimum
radius that a bubble must have in order to grow indefinitely,
 is given by $R_c=2 \sigma/V$,
we can rewrite the exponential in eq. (36) as
\beq
\frac{\Delta F}{{\rm T}}=\frac{F(R_c)}{{\rm T}}
\: \left(\frac{3 \Delta \sigma}{\sigma}
-\frac{2 \Delta V}{V}\right).
\eeq
where $\Delta \sigma$ and $\Delta V$ are the differences in surface
tension and volume energy density of the two kinds of bubbles.
The ratio $F/{\rm T}$ during the bubble nucleation was evaluated
in ref. \cite{and} in the context of the Standard Model.
In that paper it was found that $F/{\rm T}$ varies from $ \sim 130$ when
the first bubble is nucleated into an horizon volume to $\sim 100$
when all the Universe has been filled up by bubbles.
The above results are stable within a few percent as the
Higgs mass is changed from 50 to 100 GeV \cite{and}
so we believe that these estimations may be valid for the MSSM as well.

Since we are supposing that inside both kinds of bubbles the condition
(22) is not
satisfied, the contribution to $\Delta F/{\rm T}$ coming from
$\Delta V/V$ is zero.
In fact, in this case the minimization condition relative to the
phase $\delta$ gives only one solution $\delta ={\cal O}(\phi_i)$
 ($i$=3,5,6,7) and
 the vacuum inside  the two kinds of bubbles is the same.
On the contrary, the surface tension contribution will be slightly different.
A convenient expression for $\sigma$ is given by \cite{and}
\beq
\sigma\approx \int_{v \left(R-\Delta R\right)}^{v
\left(R+ \Delta R\right)} \:\:\sqrt{2 V(v)} dv,
\eeq
where $\Delta R$ is the thickness of the wall.
Using the fact that inside the bubble wall $dv / d R =\sqrt{2 V}$,
$\sigma$ can be approximated by $4\bar{V}\Delta R$, where $\bar{V}$ is
the averaged value of $V$  inside the bubble wall. If, for instance, we
take $\phi_{A}\ll\phi_B$ (this assumption is not restrictive)
a straightforward calculation gives
\begin{eqnarray}
\frac{\Delta F}{{\rm T}}&=&\frac{F}{{\rm T}}\:\frac{\Delta F}{F}=3\:\frac{F}{
{\rm T}}
\frac{\Delta \sigma}{\sigma},\nonumber\\
&\simeq& 10\:\frac{F}{{\rm T}}\:\frac{m_A^2}{{\rm T}_2^2}
\frac{\cos^2\beta}{h_t^2}\:\phi_B,
\end{eqnarray}
where $m_A$ is the pseudoscalar mass, ${\rm T}_2\sim 100 \:{\rm GeV}$ is the
temperature at which the effective potential develops a flat direction
in the origin, and $h_t$ is the top Yukawa coupling.

The relative abundance of one kind of bubbles on the other one will give
a global baryon number density to entropy ratio
\beq
B=\left(\frac{n_{B}}{s}\right)_{local}\frac{N_{+}-N_{-}}{N_{+}+N_{-}}
\simeq  \left(\frac{n_{B}}{s}\right)_{local}\:\frac{\Delta F}{{\rm T}},
\end{equation}
where $\left(n_{B}/s\right)_{local}$ is given in eq. (30).

Remembering that $F/{\rm T}$ is ${\cal O}(100)$, a value of $B$ of order
$10^{-11}$ \cite{wal} can be achieved with a phase $\phi_{B}\simeq
10^{-5}-10^{-6}$. Such a small phase  makes the supersymmetric contribution
to the {\cal CP}-violation phenomenology unobservable. Moreover, this value
is much smaller than that required in previous analysis on supersymmetric
spontaneous baryogenesis which makes use of explicit phases as the only
source of {\cal CP}-violation \cite{cohen}, whereas in our mechanism explicit
phases are only needed to achieve an asymmetry in the number of
`plus' and `minus' bubbles, while the dominant source of {\cal CP}-violation is
the
spontaneous one inside the bubble wall.

Values for the
`explicit' phases as small as $10^{-5}-10^{-6}$ might be generated through
the RGE's due to the complexity of the Yukawa couplings \cite{dug} even if
the soft SUSY breaking parameters  are chosen to be real at the
SUSY breaking scale as suggested by the simplest class of supergravity
theories \cite{ibanez}.
\vspace{1. cm}
\\
{\large {\bf 4. Conclusions}}

\vskip 0.8 cm

In the present paper we have investigated the role of spontaneous CP breaking
at finite temperature in the MSSM  \cite{cp} for the generation of the baryon
asymmetry at the electroweak scale. We have shown that a space-time dependent
relative phase between the two Higgs fields may act as a charge potential for
the baryon number, along the lines of the spontaneous baryogenesis mechanism
\cite{cohen,nel}. This scenario requires a sufficiently strong first order
EWPT, proceeding via bubble nucleation. This condition seems to be very
critical in the MSSM, however a thorough analysis covering all the parameter
space is still needed. In particular, in the case of a light pseudoscalar
mass, which is relevant for SCPV, the two masses of the neutral scalar
Higgses are not very different, and it is not possible to reduce the
minimization of the scalar potential to a one-dimensional problem, as it
is usually done.

A further source of uncertainty of the present results comes from the
evaluation
of the rate of B-violating anomalous processes inside the bubble walls,
where spontaneous baryogenesis is effective, and of the thickness of the walls
themselves. It seems that in the MSSM, the walls are sufficiently thick to
ensure equilibrium conditions for sphaleron-like transitions. We believe
that these uncertainties might lead to an overestimation of the BAU by
roughly one order of magnitude.

In general, the extra CP violation in the MSSM relative to the SM gives
rise to fine tuning
problems. Namely, the experimental bound on the neutron EDM, implies that
the `explicit' phases must be smaller than $\sim 10^{-3}$, instead of their
`natural' values $\phi\sim 1$. In the case of spontaneous CP violation  {\it
at zero temperature}, the fine tuning is even stronger: in order to have a
`spontaneous' phase $\delta\sim 10^{-3}$ one must require that the ratio
in eq. (5) differs from unity by less than $10^{-6}$ \cite{kane}.
In the scenario we have investigated the fine tuning is not so stringent.
Indeed, we have found that in order to generate the observed amount of $B$,
the `explicit' phases may be as small as $10^{-5}$. Such phases may naturally
be
induced by the Yukawa couplings through the RGE's if the soft SUSY breaking
parameters are real at the SUSY breaking scale, as predicted by the simplest
class of supergravity theories \cite{ibanez}.
Moreover, since the `spontaneous' phase
$\delta$ disappears as ${\rm T}$ goes to zero, it does not
receive any constraint
from the neutron EDM. On the other hand, as we have discussed in Sect.
{\bf 3.2}, the requirement that CP is violated inside the bubble walls at
${\rm T}\sim$100 GeV, constrains $\overline{m}_3^2$ to be less than
$\sim \lambda_{6, 7}{\rm T}^2$. This condition is able to fix the mass
of the pseudoscalar with an uncertainty of order of 10 GeV, as is shown in
fig. 1, so the fine tuning needed in the present scenario is only at the
$10^{-1}$ level.

The upper bounds on the pseudoscalar mass shown in fig. 2 are well below
100 GeV, so
the next experimental results from LEP II will be able to rule out the
possibility of generating the BAU through spontaneous CP violation at the
electroweak scale.
\vspace{1. cm}\\

\centerline{\bf Acknowledgments}

It is a pleasure to express our gratitude to G. Dvali, R. Garisto, G.F.
Giudice,
A. Masiero, G. Senjanovi\'c and F. Zwirner, for many stimulating discussions.
We also thank the CERN Theory Division, where this work was completed, for the
kind hospitality.
\vskip 2.cm
%
\def\MPL #1 #2 #3 {Mod.~Phys.~Lett.~{\bf#1}\ (#3) #2}
\def\NPB #1 #2 #3 {Nucl.~Phys.~{\bf#1}\ (#3) #2}
\def\PLB #1 #2 #3 {Phys.~Lett.~{\bf#1}\ (#3) #2}
\def\PR #1 #2 #3 {Phys.~Rep.~{\bf#1}\ (#3) #2}
\def\PRD #1 #2 #3 {Phys.~Rev.~{\bf#1}\ (#3) #2}
\def\PRL #1 #2 #3 {Phys.~Rev.~Lett.~{\bf#1}\ (#3) #2}
\def\RMP #1 #2 #3 {Rev.~Mod.~Phys.~{\bf#1}\ (#3) #2}
\def\ZP #1 #2 #3 {Z.~Phys.~{\bf#1}\ (#3) #2}

\newpage

\noindent{\bf Figure Caption}
\begin{itemize}
\item[{\bf Fig. 1}]{Upper and lower bounds for the pseudoscalar mass
versus $\tan\beta$ coming from the requirement that $0<v_{max}/{\rm
T}\simless 1$, see text. We have fixed  $\tilde{m}_Q=350$ GeV, $\mu$=200 GeV,
$A_t$=50 GeV, and $m_t$=130 GeV.}
\item[{\bf Fig. 2}]{Absolute upper and lower bounds on the pseudoscalar
mass when all the other free parameters vary in their allowed ranges
(solid lines). The experimental lower bound (dashed line) is also shown.}
\end{itemize}

\end{document}